\documentstyle[twoside,fleqn,espcrc2]{article}

\def\GSW_sign{}
\newcommand{\ba}{\begin{array}}
\newcommand{\ea}{\end{array}}
\newcommand{\beq}{\begin{equation}}
\newcommand{\eeq}{\end{equation}}
\newcommand{\be}{\begin{equation}}
\newcommand{\ee}{\end{equation}}
\newcommand{\bea}{\begin{eqnarray}}
\newcommand{\eea}{\end{eqnarray}}

\def\bra{\langle}
\def\ket{\rangle}

\def\a{\alpha}
\def\b{\beta}
\def\g{\gamma}

\def\l{\lambda}
\def\m{\mu}
\def\n{\nu}
\def\G{\Gamma}

\def\to{\rightarrow}

\def\npb#1#2#3{    {\it Nucl. Phys. }{\bf B\,#1} (19#2) #3}
\def\plb#1#2#3{    {\it Phys. Lett. }{\bf B\,#1} (19#2) #3}
\def\prd#1#2#3{    {\it Phys. Rev. }{\bf D\,#1} (19#2) #3}

\def\prl#1#2#3{    {\it Phys. Rev. Lett. }{\bf #1} (19#2) #3}

\newcommand{\AmS}{{\protect\the\textfont2
  A\kern-.1667em\lower.5ex\hbox{M}\kern-.125emS}}

\title{
\phantom{a}     \hfill        {\small BUTP--98/25} \\   
\vspace{-0.8cm}
\phantom{a}     \hfill        {\small MPI/PhT--98--69}\\      
\vspace{-0.8cm}
\phantom{a}     \hfill        {\small hep-ph/9809468}\\[3em]       
Radiative Corrections in Inclusive Rare B Decays
\thanks{based on an invited talk 
given by T.H. at 
        the International Euroconference on Quantum Chromodynamics (QCD 98),
        Montpellier, France, 2-8 July 1998.} }

\author{Christoph Greub\address{Institute of Theoretical Physics, University of 
Bern\\
                             Sidlerstrasse 5, CH-3012 Bern, Switzerland}%
               and
        Tobias Hurth\address{Max-Planck-Institute for Physics, 
Werner-Heisenberg-Institute\\
        F\"ohringer Ring 6, D-80805 Munich, Germany}}

\begin{document}
\vspace{-1cm}
\begin{abstract} 
We report recent theoretical progress in the analysis of radiative corrections 
in inclusive
rare
B-decays. In view of the B-factories, currently under construction at SLAC and 
KEK, and 
of the upgraded CESR experiment, the experimental status of rare B decays is
expected to improve significantly in the near future.
We review the complete NLL QCD calculations 
of the inclusive $b \rightarrow s \gamma$ and the 
$b \rightarrow d  \gamma$ decay rates. We also 
discuss  recently computed electroweak corrections  and further
improvements which lead to the current theoretical prediction of 
the $b \rightarrow s \gamma$ decay rate of 
${\cal B}(B \to X_s \gamma) = (3.32 \pm 0.30)
\times 10^{-4}$.
We shortly comment on the theoretical uncertainty and on implications 
to  physics beyond the SM. We collect the experimental data already available
from CESR and LEP and discuss experimental and theoretical problems
regarding the photon energy spectrum.
\end{abstract}

\maketitle

\section{Introduction}
In the Standard model (SM) rare B meson decays
%Among rare B decays, the $b \rightarrow s \gamma$ decay is the only
%mode which is already measured. The special interest in measuring 
%these rare B decay modes is twofold:  Rare B-decays 
%directly probe  the  Standard Model (SM) on the quantum level;
%as any Flavour Changing Neutral Current process, they do not arise 
%at the tree-level in the SM but 
are  induced 
by one-loop W  exchange diagrams; therefore, new contributions 
where some of the SM particles in the loop  are  
replaced by nonstandard particles
like a charged Higgs boson, a gluino or a chargino,  are not suppressed 
by an extra factor $\alpha/4\pi$ relative to the standard model amplitude. 
The resulting sensitivity for nonstandard contributions implies the possibility
for an indirect observation of new physics, a strategy complementary to the
direct production of new particles.
The $B \to X_s \gamma$ decay, for example, plays 
an important role in  restricting the parameter space of 
extensions  of the SM like the minimal supersymmetric standard
model (MSSM) \cite{Susy,SusySusy},
in spite of the fact that the accuracy of the
experimental data on  ${\cal B}(B \to X_s \gamma)$   
used for such analyses 
is not better than $30 \%$ at present. In this respect, also the $B \rightarrow 
X_d \gamma$ decay
is of specific interest because its Cabibbo-Kobayashi-Maskawa (CKM) 
suppression by the factor $|V_{td}|^2/|V_{ts}|^2$ in the SM may 
not be true in extended models.

Within the framework of the SM, rare B decays are important   
for constraining  the CKM matrix elements:
For example the $b \rightarrow s \gamma$ mode can be used to extract
$|V_{ts}|$. The is possibly the most direct measurement of this CKM matrix 
element, as the decay mode $t \to W^+ s$ is difficult to measure.
%which is not well accessible by measuring the direct
%decays of the top quark like $t \rightarrow W^+ s$. 
%The present known experimental measurement of the inclusive $b \rightarrow 
%s \gamma$ mode provides us with very important information on $|V_{ts}|$. 
Analogously, a future measurement of 
the $B \rightarrow X_d \gamma$ decay rate will help to drastically reduce
the presently allowed region
% allow for a theoretically 
%reliable determination 
of the CKM-Wolfenstein parameters $\rho$ and $\eta$.

%For both reasons, precise experimental and theoretical work
%on these decay modes is required. 
%
In contrast to exclusive 
decay channels, 
inclusive decay modes are theoretically clean in the sense 
that no specific model is needed
to describe the final hadronic state. 
%which, however, are experimentally better accessible.
Nonperturbative effects in the inclusive modes are well 
under control due to heavy quark effective theory. 
For example, the decay width $\G(B \to X_s \gamma)$ 
is well approximated by the partonic decay rate
$\G(b\to X_s \gamma)$ which can be
analyzed in renormalization group improved perturbation theory. 
The class of 
non-perturbative effects which scales like $1/m_b^2$
is expected to be well below $10\%$ (\cite{Falk}, see also \cite{Alineu}). 
This numerical
statement is supposed to hold also 
for the non-perturbative contributions
which scale like $1/m_c^2$ \cite{Voloshin}.

The accuracy
in  the dominating perturbative contribution 
was recently  improved to next-to-leading precision 
\cite{AG91,Adel,Pott,GHW,Mikolaj,GH,CDGG,INFRARED}: The renormalization
scale dependence of the previous 
leading-log result at the $\pm 25\%$-level
was substantially reduced to $\pm 6\%$ and 
the central value increased by about $20\%$.

Much of the theoretical improvements carried out in the
context of the decay $B \rightarrow X_s \gamma$ can 
straightforwardly adapted
for the decay $B \rightarrow X_d \gamma$. Like for the
former decay, the NLL-improved and
power-corrected decay rate for $B \rightarrow X_d \gamma$ has
much reduced theoretical uncertainty which would allow to extract more
precisely the CKM parameters from the measured branching ratio. 

Finally, we mention that studies of direct CP violation in inclusive 
$B \rightarrow X_s \gamma$ and $B \rightarrow X_d \gamma$   decays 
in the SM and in its extensions have been presented recently 
\cite{Neubert,AG7}. 
%In the SM, the 
%asymmetry turns out to be rather small, less than $1\%$, thus 
%out of experimental reach.
%Recently, the former analyses were extended by including 
%the full NLL order information and by investigating extended models
%with enhanced chromomagnetic dipole contributions \cite{Neubert}.
%In the latter models, large CP asymmetries of $10-50\%$ are shown
%to be possible and natural; thus the observation of a sizable
%direct CP asymmtry in this inclusive decay mode would 
%be a clean signal of new physics. 
%A new analysis of direct CP violation in the SM was recently
%presented for the inclusive $B \rightarrow X_d \gamma$ mode \cite{AG7} 
%and the CP asymmetry is found to be in the range $(7-35)\%$.  

The rest of the article is organized as follows.
In section 2, we collect the experimental data recently presented by
CLEO and ALEPH and discuss the problems regarding  
the photon energy spectrum.  
In section 3, we recall the theoretical 
framework in which rare B decays can be analyzed and discuss 
the principle steps of the  complete NLL QCD calculation of the 
inclusive $b \rightarrow s \gamma$ and $b \rightarrow d  \gamma$ 
modes. 
In section 4, we review electroweak corrections 
and further refinements which lead to the current theoretical 
prediction of the $B \rightarrow X_s \gamma$ decay rate. We shortly discuss 
the theoretical uncertainty and comment on bounds on physics beyond the SM.
\section{Experimental Data from CLEO and ALEPH, Photon Spectrum}
With the B-factories, presently under construction at SLAC (Babar) and KEK
(Belle), and also with the upgraded B-factory at CESR in Cornell (CLEO III), 
the experimental situation regarding rare B decays will drastically change 
in the near future. With the expected high luminosity, radiative B decays 
will no longer be  rare events. Experimental accuracy of below $10\%$ 
in the inclusive $b \rightarrow s \gamma$ mode appears to be possible and even 
the measurement of exclusive  $b \rightarrow d  \gamma$ modes could be in 
reach. 
This was  the motivation  to increase also the accuracy of 
the theoretical prediction correspondingly.

However, experimental data is 
already available (for a review see \cite{experiment2,experiment1}):
In 1993, the first evidence for a penguin induced B meson decay was found
by the CLEO collaboration. At the CESR $e^+ e^-$ storage ring,
which operates just 
above the $B\bar B$ threshold at the $\Upsilon(4S)$ resonance, 
they measured the exclusive 
electromagnetic penguin process $B \to K^* \gamma$.
The inclusive analogue $B \to X_s \gamma$ was also found by the CLEO 
collaboration through the measurement of its characteristic photon 
energy spectrum in 1994. 
As this process is dominated by the two-body 
decay $b \to s \gamma$, its photon energy spectrum 
is expected to be a smeared delta function centered 
at $E_\gamma \approx m_b/2$, 
where the smearing is due to perturbative gluon Bremsstrahlung and due to 
the non-perturbative Fermi motion of the b quark within the B meson.
%Only the high part of the photon energy spectrum
%GeV is sensitive to the rare decay $B \to X_s \gamma$.
Some lower cutoff in the photon energy has to be imposed in order to exclude 
the background, mainly  from the non-leptonic charged 
current processes  $b \to c q \bar{q}' + \gamma$ or
$b \to u q \bar{q}' + \gamma$, which have a typical Bremsstrahlung 
spectrum being maximal at $E_\gamma =0$ and falling off for 
larger values of $E_\gamma$. 
Therefore only the ``kinematic'' branching ratio for
$B \to X_s \gamma$ in the range between 
$E_\gamma=2.2$ GeV and the kinematic endpoint at $E_\gamma= 2.7$
GeV could be measured directly.
To obtain from this measurement the total  branching ratio, one has to know
the fraction $R$ of the $B \to X_s \gamma$ events with
$E_\gamma \ge 2.2$ GeV.
This was done in \cite{AG91} where 
the Fermi motion of the b quark in the B meson was
taken into account by using the phenomenological model by Altarelli et
al. (ACCMM model) \cite{ACCMM}.
Using this theoretical input regarding the photon energy spectrum, 
the value  $R=0.87 \pm 0.06$ was used  for the fraction
by the CLEO collaboration, leading to the CLEO branching ratio
\cite{CLEOincl}
\beq 
\label{cleoincl}
{\cal B}(B \to X_s \gamma) = (2.32 \pm 0.57 \pm 0.35) \times 10^{-4} 
\eeq
Actually, there are two separate CLEO analyses. The first technique 
constructs the inclusive rate by summing up the possible
exclusive final states. Background in the
measurement of exclusive modes is naturally  low because of kinematical 
constraints 
and of the beam energy constraint.
In the second technique  one measures the inclusive photon 
spectrum near the end 
point. 
%There are very large backgrounds from the continuum, both from the 
%initial-state-radiation
%(ISR) process $e^+e^- \rightarrow q \bar q \gamma$ and from the continuum 
%reaction
%$e^+e^- \rightarrow q \bar q$. 
Background suppression is  more difficult.
For this purpose one uses topological differences between the spherical 
$B\bar{B}$ events 
and the two jets $e^+e^- \rightarrow q\bar{q}$. But the signal efficiency ($32 
\%$) is very high compared to the first technique.
The branching ratio stated above (\ref{cleoincl})  is the average 
of the two measurements, taking into account the correlation between
the two techniques. The first error is statistical and the second is systematic 
(including
model dependence). The measurement  is based on a sample of $2.2 \times 10^6 
B\bar B$
events. 

This summer (1998), CLEO has presented an improved, but preliminary 
 measurement 
\cite{CLEOneu}
which is based on $53\%$ more data ($3.3 \times 10^6$ events).
 They also used the
slightly wider $E_{\gamma}$ window starting at $2.1$ GeV. 
The relative error drops almost by a factor of $\sqrt{3}$:
\beq 
\label{cleoneu}
{\cal B}(B \to X_s \gamma) = (3.15 \pm 0.35 \pm 0.32 \pm 0.26) 
\times 10^{-4} 
\eeq
The errors represent statistics, systematics, and
the model dependence, respectively.
In ref. \cite{CLEOneu} the kinematical branching fraction
${\cal B}^{2.1}$ for photons with energies in the 
range between  2.1 GeV and 2.7 GeV is also given:
%in order to make a comparision with 
%theoretical analysis of the photon spectrum other than the 
%ACCMM model possible:
\beq 
\label{cleoneu2}
{\cal B}^{2.1}(B \to X_s \gamma) = (2.97 \pm 0.33 \pm 0.30 \pm 0.21) 
\times 10^{-4} 
\eeq
As
CLEO II still analyses more data, 
one can expect an even better measurement soon.

There is also data at the $Z^0$ peak from the LEP experiments.
The ALEPH collaboration \cite{ALEPH} has measured the
inclusive
branching ratio
\beq
\label{braleph} 
{\cal B}(H_b \to X_s \gamma) = (3.11 \pm 0.80 \pm 0.72) \times 10^{-4}. 
\eeq
It should be noted that
the branching ratio in (\ref{braleph}) involves a different weighted
average of the B mesons and $\Lambda_b$ baryons produced in $Z^0$ decays
(hence the
symbol $H_b$) than the corresponding one given by CLEO.
% which has been
%measured at  the $\Upsilon(4S)$ resonance.
High luminosity is more difficult to obtain
at higher $e^+e^-$ collision energies. Thus, $B\bar{B}$ samples
obtained by the LEP experiments are smaller than the one 
at CESR.  The rate measured by ALEPH,
is consistent with the CLEO measurement.

The uncertainty regarding the fraction $R$ of the $B \to X_s \gamma$ events with
$E_\gamma \ge 2.2$ GeV spotted in the experimental measurement should 
be 
regarded as a purely theoretical uncertainty.
As mentioned above, 
the fraction $R$ was calculated 
in \cite{AG91} using the phenomenological model by Altarelli et
al., where the Fermi motion of the b quark in the B meson 
is characterized by two parameters, the
average Fermi momentum $p_F$ of the b quark and the mass $m_q$ of the spectator 
quark.
The error on the fraction $R$ is essentially obtained by varying 
the model parameters $p_F$ and $m_q$ in the range for which the ACCMM
model correctly describes the energy spectrum of the charged lepton in
the semileptonic decays $B \to X_c \ell \nu$ and $B \to X_u \ell \nu$,
measured by CLEO and ARGUS. 
In \cite{AG91} a first comparison between the calculated photon energy 
spectrum 
and the one measured by the CLEO collaboration was presented.
% in order to
%determine the non-perturbative parameters of the model.    
The (normalized) measured
photon energy spectrum and the theoretical one are in agreement for those 
values of  $p_F$ and
$m_q$, which correctly describe the inclusive semileptonic CLEO data $B \to X_c
\ell \nu$ and $B \to X_u \ell \nu$;
at present, the data from the
radiative decays is, however, not precise enough to 
further constrain the values of $p_F$ and $m_q$. 
The best fit between 
the theoretical and measured photon energy
spectrum is obtained for $p_F=450$ MeV and $m_q=0$.
%One should mention that that  the analysis \cite{AG91} on the
%photon energy spectrum, in particular 
%the calculation of the fraction $R$ in the 
%ACCm model used by CLEO, does not included the full
%NLL information which is alreday available.

Besides this phenomenological model by Altarelli et al.,
 more fundamental theoretical methods 
are  available today to implement the bound state effects, namely by making use 
of operator product expansion techniques in the framework of heavy quark 
effective theory (HQET).
A new analysis along these lines  was recently presented \cite{Kagan}.
Unfortunately, the operator product expansion breaks down 
near the endpoint of the photon energy  spectrum and therefore an infinite 
number of leading-twist corrections  has to be resummed into
a non-perturbative universal "shape function" which determines the
light-cone momentum distribution of the b-quark in the B meson. The physical 
decay distributions are then obtained from a convolution of parton model spectra 
with this shape function.
At present this function cannot be calculated, but there is at least some 
information 
on the moments of the shape function which are related to the forward matrix 
elements 
of local operators. Ans\"atze  for the shape 
function, constrained by the latter information, are used. 
In contrast to the older analysis based on the ACCMM model, 
the new analysis of Kagan and Neubert \cite{Kagan}
includes the full NLL information. Their  
fraction
$R=0.78^{+0.09}_{-0.11}$ (for the energy cut $E_\gamma > 2.2$ GeV) 
is significantly 
smaller than the factor used by CLEO. 
The larger error on $R$ implies that the theoretical uncertainty in the 
calculation of Fermi motion effects has been underestimated until now.
Clearly, a lower experimental cut decreases the sensitivity 
to the parameters of the shape function (or, more general, the 
model dependence as one can already see from the new CLEO measurement.)
Another future aim should be to determine the shape function
(and analogously the parameter of the ACCMM model) 
by using the high-precision measurements of the photon  energy
spectrum. 

\section{QCD Corrections at the NLL level}

Short distance QCD corrections 
enhance the partonic decay rate $ \Gamma(b \to s \g)$  by more than a factor of 
two.
These QCD effects bring in large logarithms of the form 
$\alpha_s^n(m_b) \, \log^m(m_b/M)$,
where $M=m_t$ or $M=m_W$ and $m \le n$ (with $n=0,1,2,...$).
This is a natural  feature in any process where two different mass scales
are present.
In order to get a reasonable result, one has  to resum at least
the leading-log (LL) series ($m=n$).  
Working to next-to-leading-log (NLL) precision means that one is also resumming 
all the
terms of the form $\a_s(m_b) \, \left(\a_s^n(m_b) \, \ln^n (m_b/M)\right)$.

The error of the leading logarithmic (LL) result \cite{counterterm}
was  dominated by a large renormalization scale dependence 
at the $\pm 25\%$ level which indicates the importance of the NLL series.
Moreover, such a NLL program 
is also important in order to ensure validity
of renormalization group improved perturbation theory in the example  under 
question.
It was recently found in Multi-Higgs Doublet Models \cite{BG} 
that the truncation of the perturbative series at the NLL level is often not
appropriate.

A suitable framework to achieve the necessary resummations 
of the large logs is an  effective 
low-energy theory, obtained by integrating out the
heavy particles which in the SM are the top quark and the $W$-boson. 
The effective Hamiltonian relevant for $b \to s \gamma$ and
$b \to s g$ in the SM and many of its extensions reads 
\begin{equation}
\label{heff}
H_{eff}(b \to s \gamma)
       = - \frac{4 G_{F}}{\sqrt{2}} \, \lambda_{t} \, \sum_{i=1}^{8}
C_{i}(\mu) \, O_i(\mu) \quad 
\end{equation}
where $O_i(\m)$ are the relevant operators,
$C_{i}(\mu)$ are the corresponding Wilson coefficients,
which contain the complete top- and W- mass dependence,
and $\lambda_t=V_{tb}V_{ts}^*$ with $V_{ij}$ being the
CKM matrix elements. The CKM dependence globally factorizes,
because we work in the approximation $\l_u=0$ (for $B \to X_s \gamma$).

Neglecting operators with dimension $>6$ which are suppressed 
by higher powers of $1/m_{W/t}$ and using the equations
of motion for the operators, one arrives at the following basis 
of dimension 6 operators 
\bea
\label{operators}
O_1 &=& \left( \bar{c}_{L \b} \g^\m b_{L \a} \right) \,
        \left( \bar{s}_{L \a} \g_\m c_{L \b} \right)\,, \nonumber \\
O_2 &=& \left( \bar{c}_{L \a} \g^\m b_{L \a} \right) \,
        \left( \bar{s}_{L \b} \g_\m c_{L \b} \right) \,,\nonumber \\
O_7 &=& \frac{e}{16\pi^{2}} \, \bar{s}_{\a} \, \sigma^{\m \n}
      \, (m_{b}(\mu)  R) \, b_{\a} \ F_{\m \n} \,,
        \nonumber \\
O_8 &=& \frac{g_s}{16\pi^{2}} \, \bar{s}_{\a} \, \sigma^{\m \n}
      \, (m_{b}(\mu)  R) \, \frac{\l^A_{\a \b}}{2} \,b_{\b}
      \ G^A_{\m \n} \quad .
\eea

Because the Wilson coefficients of the penguin induced Four-Fermi
operators $O_3,..O_6$ are very small, we do not list them here.

In this framework the next-to-leading logarithmic terms 
$\a_s(m_b) \, (\a_s^n(m_b) \log^n(m_b/m_{W/t}))$ 
in the $b \to s \gamma$ amplitude
have two sources:\\ 
{\bf 1} $\bullet$ The corrections to the Wilson coefficients $C_i(\mu)$ at the 
scale $\mu \approx m_b$.\\
 {\bf 2} $\bullet$ The corrections to the matrix elements of the operators $O_i$ 
also at the low-energy scale $\mu \approx m_b$.\\
Only the sum of the two 
contributions is renormalization scheme 
independent and in fact, 
from the $\mu$-independence of the effective Hamiltonian,
one can derive a renormalization group equation 
(RGE) for the Wilson 
coefficients $C_i(\mu)$:
\be
\label{RGE}
\mu \frac{d}{d\mu} C_i(\mu) = \gamma_{ji} \, C_j(\mu) \quad ,
\ee  
where the $(8 \times 8)$ matrix $\gamma$ is the anomalous dimension
matrix of the operators $O_i$.
The standard procedure to calculate the two contributions involves the
following three steps:\\
{\bf ad 1a} $\bullet$  One matches the full standard model theory
with the effective theory at the scale $\m=\m_W$, where
$\m_W$ denotes a scale of order $m_W$ or $m_t$. At this scale,
the matrix elements of the operators  in the 
effective theory lead to the  same logarithms  as the full theory
calculation. 
Consequently, the Wilson coefficients 
$C_i(\m_W)$ only pick up small QCD corrections,
which can be calculated in fixed-order perturbation theory.
In the LL (NLL) program, the matching has to be worked out to order
$\a_s^0$ ($\a_s^1$).\\
{\bf ad 1b} $\bullet$ Solving the RGE (\ref{RGE}) and using the $C_i(\m_W)$ 
of Step 1a as initial conditions, one performs the   
evolution of these Wilson coefficients from 
$\m=\m_W$ down to $\m = \m_b$, where $\m_b$ is of the order of $m_b$.
As the matrix elements of the operators evaluated at the low scale
$\m_b$ are free of large logarithms, the latter are contained in resummed
form in the Wilson coefficients. For a LL (NLL) calculation, this RGE step
has to be performed using the anomalous dimension matrix $\gamma_{ji}$  up 
to order $\a_s^1$ ($\a_s^2$).\\
{\bf ad 2} $\bullet$ The corrections to the matrix elements 
of the operators $\bra s \g |O_i (\mu)|b \ket$ at the scale  $\mu = \m_b$
have to be calculated to order $\a_s^0$ ($\a_s^1$) in the LL (NLL)
calculation.

%This ambitious NLL enterprise was completed now. 
All three steps (1a,1b,2) to NLL precision involve rather difficult
calculations.   The most difficult part in Step 1a is the 
 two-loop (or  order $\a_s$) matching of the dipole operators $O_7$ and $O_8$. 
It involves two-loop diagrams both in the full and in the effective theory. 
It was worked out by Adel and Yao \cite{Adel} some time ago. 
As this is a crucial step in the NLL program,
Greub and Hurth confirmed their findings in a detailed re-calculation 
using a different method \cite{GH}.
Recently, two further recalculations of this result were presented
\cite{CDGG,INFRARED}.
Step 2 basically consists of Bremsstrahlung corrections and virtual
corrections. While the Bremsstrahlung corrections
were worked out some time ago by Ali and Greub \cite{AG91} and have
 been confirmed and extended by Pott \cite{Pott}. A  
complete analysis of the virtual two-loop corrections (up to the contributions 
of the Four-Fermi operators with very small coefficients) was presented
by Greub, Hurth and Wyler \cite{GHW}. 
%The latter  calculation involves two-
%loop diagrams where the full charm dependence has to be taken into account.   
%By using Mellin-Barnes techniques in the Feynman parameter integrals, 
%the result of these two-loop
%diagrams was obtained in the form
%\be
%\label{Mellin}
%c_0 + \sum_{n=0,1,2,...;m=0,1,2,3} c_{nm} \left( \frac{m_c^2}{m_b^2}
%\right)^n \, \log^m \frac{m_c^2}{m_b^2} 
%\ee
%where the quantities $c_0$ and $c_{nm}$ are independent of $m_c$.
%Note, that a finite result is obtained in the limit $m_c \to 0$,
%as there is no naked logarithm of $m_c^2/m_b^2$. This observation
%is of some
%importance in the $b \to d \gamma$ process, where the $u$-quark
%propagation
%in the loop is not CKM suppressed (see below).
%The main result of this analysis
%consists in a drastic reduction of the renormalization 
%scale uncertainty from about $\pm 25\%$ to about $\pm 6\%$.
%Moreover, the central value was shifted outside the $1\sigma$ bound of the
%former CLEO measurement (\ref{cleoincl}).
%However, at that time, the essential
%coefficient $C_7(\m_b)$ was only known to leading-log precision. 
%It was therefore
%unclear how much the overall normalization will be changed when
%the NLL value for $C_7(\m_b)$ is used. 
The order $\a_s^2$ anomalous matrix (Step 1b) has been 
worked out
by Chetyrkin, Misiak and M\"unz \cite{Mikolaj}. The extraction of some of 
the elements in the $O(\a_s^2)$ anomalous
dimension matrix involves pole parts of three-loop diagrams. 

Combining the NLL calculations of all the three steps (1a+b,2), 
the first theoretical prediction to NLL  precision 
for the branching ratio 
was presented in \cite{Mikolaj}:
\be
\label{mikolajend}
BR(B \to X_s \g)=(3.28 \pm 0.33) \times 10^{-4}.
\ee
The theoretical error 
has two dominant sources:  The 
$\mu$ dependence is reduced to about $6\%$. The other main uncertainty of $5\%$ 
stems from the $m_c/m_b$ dependence. This first theoretical NLL prediction already included the 
nonperturbative correction which scale with $1/m_b^2$ which are rather small
at the $1\%$ level.
%As also the semileptonic decay rate which is usually used
%to normalize the $B \rightarrow X_s \gamma$ decay rate 
%has also $1/m_b^2$ corrections which are negative (see e.g. \cite{Manohar}) 
%these non-perturbative corrections tend to cancel
%in the ratio ${\cal B}(B \to X_s \gamma)/{\cal B}(B \to X_c e \bar{\nu_e}$ 
%and only about $1\%$ remains.
Later, also nonperturbative contributions from $c\bar c$ intermediate states 
were considered which scale with $1/m_c^2$ \cite{Voloshin}.  
%As the expansion parameter is $m_b \Lambda_{QCD}/m_c^2 \approx 0.6$
%(rather than $\Lambda^2_{QCD}/m_c^2$), it was not a priori clear
%whether formally higher order terms in the $m_c$ expansion are
%numerically suppressed. 
Detailed investigations \cite{Voloshin}
show that 
these contributions enlarge
the branching ratio by $3\%$.

The $B \rightarrow X_d \gamma$ decay can be treated analogously.
The effective Hamiltonian  is the same 
in  the processes $b \to s \gamma (+g)$ and $b \to d \gamma (+g)$
up to the obvious
replacement of the $s$-quark field by the $d$-quark field.  
However, as $\lambda_u$ for $b \to d \gamma$ is not small relative to
$\lambda_t$ and $\lambda_c$, one also has to encounter 
the operators proportional
to $\lambda_u$.
The matching conditions $C_i(m_W)$
and the solutions
of the RG equations, yielding $C_i(\mu_b)$, coincide
with those needed for the process $b \to s \gamma (+g)$.
The power corrections in $1/m_b^2$ and $1/m_c^2$ (besides the CKM factors)
are also the same for both modes. However, the so-called long-distance 
contributions
from the intermediate $u$-quark in the penguin loops are different.
These are  suppressed in the $B \rightarrow X_s \gamma$ mode due to the 
unfavorable CKM matrix elements. In $B \rightarrow X_d \gamma$, however, there 
is 
no CKM-suppression  and one
has to include the long-distance intermediate $u$-quark contributions. 
The non-perturbative contribution generated by the $u$-quark loop
can only be modeled at present.
%\footnote{ It
%must be stressed that
%there is no spurious enhancement  of the form $\ln (m_u/\mu_b)$ 
%in the perturbative contribution to the matrix elements
%$<\overline{X_d}\gamma|O_{iu}|\overline{B}>$ ($i=1,2$)
%as shown by the explicit calculation in \cite{GHW} and also
%discussed more recently in \cite{STERMAN}. In other words, the limit $m_u
%\to 0$ can be taken safely.}
\section{Theoretical Predictions, Current Status}
The prediction for the partonic $b \rightarrow s \gamma$ decay rate  is usually 
normalized by the semileptonic decay rate in order to get rid of
uncertainties 
related to the fifth power of the b quark mass.
Moreover, often an explicit lower cut on the photon energy in the
Bremsstrahlung correction is made: 

\beq
R_{quark}(\delta) = 
\frac{\Gamma[ b \to s \gamma]+\Gamma[ b \to s \gamma  gluon]_\delta}{\Gamma[ b 
\to X_c e \bar{\nu}_e ]}
\eeq
where the subscript $\delta$ means that only photons with energy 
$E_{\gamma} > (1-\delta) E_{\gamma}^{max} = (1-\delta) \frac{m_b}{2}$ are 
counted. 
The ratio $R_{quark}$ is divergent in the limit $\delta \rightarrow 1$ due to 
the 
unphysical soft photon divergence in the subprocess $b \rightarrow s \gamma 
gluon$. 
In this limit only the sum of $\Gamma[b \to s \gamma]$,
$\Gamma[b \to s  gluon]$ and $\Gamma[b \to s \gamma gluon]$ is a reasonable 
physical quantity, in which all divergences cancel out. 
In \cite{GHW} the limit $\delta \rightarrow 1$ is taken and the singularities
are removed by adding the virtual photon corrections to $b \rightarrow s gluon$.
In  \cite{Mikolaj} the "total" partonic $b \rightarrow s \gamma$ is defined 
by using the value at $\delta=0.99$. However, in \cite{Kagan} it was shown that 
this
is not the most suitable choice, because the theoretical result becomes 
more sensitive to the unphysical soft-photon divergence than the first 
numerical results indicated. 
The authors of \cite{Kagan} use $\delta=0.90$ as their optimized definition 
of the total decay rate. They  also
suggest to 
%give up the concept of a "total" decay rate and 
directly compare theory and experiment 
using the same energy cuts as CLEO($E_{\gamma}>2.1 (2.2)$ GeV). Then   
the theoretical uncertainty  regarding the photon energy spectrum discussed in 
section 2 would occur 
naturally in the theoretical prediction. 

Using the measured semileptonic branching ratio ${\cal B}^{sl}_{exp.}$,
the  branching ratio ${\cal B}(B \to X_s \gamma)$ is given by 
\beq
{\cal B}(B \to X_s \gamma) = R_{quark}  \times {\cal B}^{sl}_{exp.} (1 + 
\Delta_{nonpert})
\eeq 
where 
$\Delta_{nonpert}$ contains 
the $1/m^2_b$ and $1/m^2_c$ corrections, whose effects
amount to $+1\%$ and $+3\%$, respectively.
%summed in 
%have a numerical effect of $+1\%$ respectively 
%$+3\%$ on the branching ratio (see last section).

The NLL QCD analysis for the branching ratio 
${\cal B}(B \to X_s \gamma)$
is 
further based on the following values for the input parameters \cite{BG}:
$\alpha_s(M_Z)=0.119\pm 0.004$, $m_b-m_c=3.39\pm 0.04$ GeV,
$m_c/m_b=0.29\pm 0.02$, 
$ m_t(\mbox{pole})=(175\pm 5)$~GeV, ${\cal B}_{sl}=(10.49 \pm 0.46)\%$,
$\alpha_{em}^{-1}=(130.3 \pm 2.3)$. 
For the CKM  matrix factor $|V_{ts}^*V_{tb}/V_{cb}|^2$ the number $(0.95 \pm 
0.03)$ 
is used.
For these values of the input parameters, one gets
%the theoretical branching ratio for the decay  $B \rightarrow X_s \gamma$
%in the SM is
\beq
\label{AG7end}
{\cal B}(B \rightarrow X_s \gamma) = (3.57 \pm 0.32)\times 10^{-4}. 
\eeq
The shift in the central value compared  
with the prediction (\ref{mikolajend}) 
has different sources. 
The recently discovered nonperturbative corrections  scaling with $m_c^2$ 
are included in addition.
Besides slightly different input parameters,
the discussed convention $\delta \rightarrow 1$ is used (\ref{AG7end}),
 while in 
(\ref{mikolajend}) $\delta=0.99$ was chosen.
The central value is also sensitive to the details how the uncalculated  
next-next-leading 
order terms are discarded. We note that in (\ref{AG7end})
the factor $1/\Gamma_{sl}$ is not expanded in $\alpha_s$. (If one does so, the 
central value
in (\ref{AG7end}) decreases from $3.57 \times 10^{-4}$ to 
$3.46 \times 10^{-4}$.) 

We emphasize that 
for a comparison with the ALEPH measurement (\ref{braleph}) one should use 
consistently
the measured semileptonic branching ratio ${\cal B}(H_b \to X_{c,u} \ell 
\nu)=(11.16 \pm 0.20)\%$.
This leads to a larger theoretical prediction for the LEP experiments.

% The dependence on the renormalization scale $\mu_b$ is 
% obtained
% by the variation $m_b/2 < \mu_b < 2 m_b$. As mentioned above, this
% scale dependence is substantially reduced  from about 
% $\pm 25\%$ (LL) to about $\pm 6\%$ (NLL) in the SM prediction.
% To produce the central value of the NLL result out of the 
% LL result the scale has to be chosen to be around $\mu_b=2.1$ GeV.
% This suggests that this  convention tends to underestimate
% the size of the next-order contribution at least in this example.
%

Quite recently, A. Czarnecki and W. Marciano \cite{Marciano} 
calculated part of the electroweak two-loop contributions, namely contributions
from  fermion loops in gauge boson
propagators ($\gamma$ and $W$) and from short-distance photonic loop corrections
which are considered to be the two dominant classes of electroweak
corrections. They  found that these new contributions reduce the $R_{quark}$ 
(\ref{AG7end}) by $9\%$. 
They observed   that the on-shell value of the fine
structure constant  $1/\a_{em}=137$ is more appropriate for real
photon emission instead of
the value $1/\a_{em} = (130.3 \pm 2.3)$ used in previous analyses. 
Their loop calculations confirmed this expectation.
This change in $\a_{em}$ represents the main 
reduction of $-5\%$ in $R_{quark}$.
Also recently, Strumia \cite{Strumia} made a complete analysis of the 
heavy top and the heavy Higgs corrections 
to $b \rightarrow s \gamma$ in the limit
$m_W \rightarrow 0$. 
The correction is below $1\%$ which indicates  that
the $-2.2\%$ correction from the fermion loop contribution stated in 
\cite{Marciano} could be an overestimation. 
Kagan and Neubert \cite{Kagan} improved the QED analysis made in 
\cite{Marciano} 
by resumming the contributions of order 
$\alpha ln(m_W/\mu_b) (\alpha_s ln(m_W/\mu_b)^n$
to all orders (while in \cite{Marciano} only the $n=0$ contribution
was included). 
This resummation decreases the QED corrections.

Including only the resummed QED corrections, using the on-shell value of 
$\alpha_{em}$
and 
working with the 
convention $\delta \rightarrow 1$ in $R_{quark}$, we end up with 
the current theoretical prediction 
\beq
\label{end2}
{\cal B}(B \rightarrow X_s \gamma) = (3.32 \pm 0.14 \pm 0.26)\times 10^{-4},
\eeq
where the first error represents the uncertainty regarding the scale dependences
on $\mu_b$ and $\mu_W$, while the second error is the
uncertainty due to the input parameters. 

A complete NLL calculation
of the $B \rightarrow X_s \gamma$ branching ratio in the simplest
extension of the SM, namely the Two-Higgs-Doublet Model (2HDM), was recently
presented in \cite{CDGG,BG}. In the 2HDM of Type II (which already represents 
a good approximation for 
gauge-mediated
supersymmetric models with large $\tan\beta$ where the charged Higgs
contribution dominates the chargino contribution) a lower
bound on the mass of the charged Higgs boson of about $250$ GeV was found
for large  $\tan\beta$. This bound was based on the CLEO upper limit 
($95 \%$ C.L.) of 
$4.2 \times 10^{-4}$ for ${\cal B}(B \to X_s \gamma)$ \cite{CLEOincl}
and the electromagnetic corrections discussed above were not included.
Taking into account these corrections, and using the most recent
upper CLEO bound ${\cal B}(B \to X_s \gamma) < 4.5 \times 10^{-4}$ (at
$95 \%$ C.L.) \cite{CLEOneu}, the lower bound for the mass 
of the charged Higgs bosons
is at about 165 GeV \cite{newcontour} for large $\tan \beta$.
This value should be compared 
with the lower bound of 45 GeV found in the direct search for the charged 
Higgs boson at LEP-I. The validity of the bound at $\sim 55$ GeV from
LEP-II was criticized in \cite{Djouadi}. 
Quite recently, 
a more general SUSY scenario was presented 
\cite{susyneu}, where
in particular the possibility of destructive interference of the chargino 
and the charged Higgs contribution is analyzed. 

Instead of making a theoretical standard model 
prediction for the branching ratio
${\cal B}(B \to X_s \gamma)$,
one can use the experimental data  and theory in order to directly determine
the combination $|V_{tb} V_{ts}^*|/|V_{cb}|$ of CKM matrix elements; 
in turn, one can determine 
$|V_{ts}|$, by making use of the relatively well  known
CKM matrix elements $V_{cb}$ and $V_{tb}$. 
An update of the analysis done by A. Ali \cite{aliokt97}, 
using the CLEO data (\ref{cleoneu}),
the ALEPH data (\ref{braleph}), and including
all new contributions on the theoretical side, leads to
\begin{eqnarray*}
\frac{|V_{ts}^* V_{tb}|}{|V_{cb}|} &=& 0.95 \pm 0.08_{exp.} \pm 0.05_{th.}
\quad \rm{CLEO} \nonumber \\
\frac{|V_{ts}^* V_{tb}|}{|V_{cb}|} &=& 0.91 \pm 0.15_{exp.} \pm 0.04_{th.}    
\quad \rm{ALEPH.} 
\nonumber
\end{eqnarray*}
The average of the two measurements yields
\beq
 \frac{|V_{ts}^* V_{tb}|}{|V_{cb}|} = 0.93 \pm 0.09  \pm 0.03
=0.93 \pm 0.10
\eeq
where in the very last step the theoretical and experimental errors were
added in quadrature.  
Using
$|V_{tb}|=0.99 \pm 0.15$ from the CDF 
measurement and  $|V_{cb}|=0.0393 \pm 0.0028$  
extracted from semileptonic $B$ decays 
, one obtains
\beq
\label{vts}
|V_{ts}|=0.037 \pm 0.007,
\eeq
where all the errors were added in quadrature. This  is probably
the most direct determination of this CKM matrix element,
as mentioned in the introduction.
With an improved measurement of ${\cal B}(B \to X_s \gamma)$
and $V_{tb}$, one expects to reduce the present error on $|V_{ts}|$
by a factor of 2 or even more.

Let us finally move to the predictions for the 
$ B \rightarrow X_d \gamma$ decay \cite{AG7}.
One finds that for $\mu_b=2.5$ GeV (and the central values of the input
 parameters) the difference between the LL and NLL results is 
$\sim 10\%$, increasing the branching ratio in the NLL case.
For a fixed value of the CKM-Wolfenstein parameters 
$\rho$ and $\eta$, the theoretical uncertainty of the branching
ratio is:
$\Delta {\cal B}(B \rightarrow X_d \gamma)/ {\cal B}(B \rightarrow X_d \gamma)   
= 
\pm (6-10)\%$. 
Of particular theoretical interest is the ratio of the
branching ratios, defined as
\begin{equation}
\label{dsgamma}
R(d\gamma/s\gamma) \equiv \frac{{\cal B}(B \to X_d \gamma)}
                           {{\cal B}(B \to X_s \gamma)},
\end{equation}
in which a good part of the theoretical uncertainties cancels. 
As expected,
the uncertainty in the ratio $R(d\gamma/s\gamma)$ is indeed smaller.
%The residual theoretical uncertainty $\Delta R/R$ is correlated with the
%value of $\rho$ and $\eta$ which represent a large, defintely 
%dominating uncertainty.
This suggests that 
a future  measurement of $R(d\gamma/s\gamma)$ will have a large impact on
the CKM phenomenology.
% which will be almost determined by experimental errors
%only.  

Varying the CKM-Wolfenstein parameters $\rho$ and $\eta$ in the range
$-0.1 \leq \rho \leq 0.4$ and $0.2 \leq \eta \leq 0.46$ and taking into
account other parametric dependences stated above, the 
results (without electroweak corrections) are
\begin{eqnarray}
\label{summarybrasy}
6.0 \times 10^{-6} &\leq &
 {\cal B}(B \rightarrow X_d \gamma)   \leq 2.6 \times 10^{-5}~, \nonumber\\
0.017 &\leq & R(d\gamma/s\gamma) \leq 0.074~.\nonumber
\end{eqnarray}
These quantities are expected to be measurable at the forthcoming high 
luminosity B facilities. At present only upper bounds on corresponding
exclusive  modes are available from CLEO II, namely ${\cal B}(B^0\to 
\rho^0\,\gamma)=<3.9\cdot10^{-5} $, ${\cal B}(B^0\to \omega\,
\gamma)=<1.3\cdot10^{-5} $, and ${\cal B}(B^-\to 
\rho^-\,\gamma)=<1.1\cdot10^{-5}$ \cite{experiment1}.

\section{Summary}
Significant theoretical progress in the analysis of 
inclusive rare radiative B-decays has been
achieved during the last few years. In particular, the completion of the
NLL QCD calculation for the dominating perturbative contribution
and further refinements on nonperturbative and  electroweak corrections 
and on the photon energy spectrum lead to the present
theoretical prediction of the $B \rightarrow X_s \gamma$ branching ratio
with a substantially  improved precision.  
These  theoretical improvements  call for more precise experimental data 
which one can expect from the B-factories and the upgraded CLEO
experiment in the next years. This experimental data will provide 
important tests of the SM and its extensions.

%%%%%% references %%%%%%%%%%%%%%%%%

\end{document}